\documentclass[11pt,draftcls,onecolumn]{IEEEtran}
\IEEEoverridecommandlockouts

\usepackage{citesort}
\usepackage[latin1]{inputenc}
\usepackage{url}
\usepackage{amsfonts}
\usepackage{amssymb}
\usepackage{amsmath}
\usepackage{theorem}
\usepackage{algorithm}
\usepackage[noend]{algorithmic}
\usepackage{multirow}
\usepackage{tikz}
\usepackage{graphicx}
\usepackage{subfigure}

\usetikzlibrary[shapes]
\usetikzlibrary{mindmap} 
\usetikzlibrary{snakes} 
\usetikzlibrary{patterns}

\newcommand{\remove}[1]{}

\title{Decentralized Architecture for Large-Scale Time-Shifted TV Systems}

\author{
  Yaning Liu and~Gwendal Simon \thanks{Yaning Liu is with INRIA Sophia Antipolis
2004, route des Lucioles -- B.P. 93
F-06902 Sophia Antipolis Cedex -- FRANCE, yaning.liu@inria.fr,
Phone:+33 4 92 38 50 16 Fax: +33 4 89 73 24 00} \thanks{Gwendal Simon is with Institut Telecom - Telecom
    Bretagne, Technopole Brest-Iroise 29238 Brest France
    gwendal.simon@telecom-bretagne.eu Phoe:+33 2 29 00 15 73 Fax:+33 2
  29 00 12 82} }

\begin{document}

\maketitle

\begin{abstract}
  An attractive new feature of connected TV systems consists in allowing users to access past portions of the TV channel. This feature, called time-shifted streaming, is now used by millions of TV viewers. We address in this paper the design of a large-scale delivery system for time-shifted streaming.  We highlight the characteristics of time-shifted streaming that prevent known video delivery systems to be used. Then, we present two proposals that meet the demand for two radically different types of TV operator. First, the Peer-Assisted Catch-Up Streaming system, namely PACUS, aims at reducing the load on the server of a large TV broadcasters without losing the control of the TV delivery. Second, the turntable structure, is an overlay of nodes that allow an independent content delivery network or a small independent TV broadcaster to ensure that all past TV programs are stored and as available as possible. We show through extensive simulations that our objectives are reached, with a reduction of up to three quarters of the traffic for PACUS and a 100\% guaranteed availability for the turntable structure. We also compare our proposals to the main previous works in the area.
\end{abstract}



\section{Introduction} \label{sec:intro}
\IEEEPARstart{F}{or} users of a\emph{ time-shifted} TV service, a
program normally broadcast at time $t$ can be viewed at any time after
$t$ (from a few seconds to many days), even if the program is still on
air. Time-shifted TV is currently commercially available through
network digital video recorders (NDVRs) and personal digital video
recorders (PDVRs). It is also available in a limited form known as
\emph{catch-up TV} where TV programs can be watched on demand after
they have been broadcast and recorded. Time-shifted TV is gaining in
popularity. According to the 2010 BARB's Thinkbox Review~\cite{barb},
time-shifted TV accounts now for 14\% of the overall TV consumption in
UK households equipped with PDVRs. It is also the TV usage that grew
at the highest rate in 2009 in the US according to Nielsen's Three
Screen Report issued in June 2010~\cite{nielsen2}. More recent reports
consistently reported that about 20 percent of all
adult US viewers have shifted more than half of their video
viewing~\cite{saymedia}.

Despite the demand for time-shifted TV services is growing fast, TV
broadcasters face cost and scalability issues with their NDVR
services, which are based on client-server architectures. First,
conventional disk-based VoD servers cannot massively ingest content,
and keep pace with the changing viewing habits of subscribers, because
they have not been designed for concurrent read and write
operations. Second, client-server delivery systems are not
cost-efficient in applications where clients require distinct portions
of a stream because they cannot use group communication techniques
such as multicast protocols. As a matter of facts, time-shifted
services managed by TV broadcasters are restricted to a time delay
ranging from one to three hours, despite only $40\%$ of users of
time-shifted TV systems (hereafter called \emph{shifters}) watch their
program less than three hours after the live program~\cite{nielsen}.

The alternative, where viewers directly use their PDVR equipment to
record the channels, has some drawbacks too. First, PDVRs can record
only a small number of programs due to their limited storage
capacity. They also require the client to know in advance the program
to record. Although the client can nowadays remotely control her
recordings, PDVRs does not offer the same flexibility as a
time-shifted service provided by a broadcaster.  Moreover, a PDVR
viewer can decide to fast forward through commercials. The question of
controlling the delivery of advertisement has become critical in the
new-generation TV world. In order to protect their advertisement
revenues, TV broadcasters need to control the content delivery and to
propose new services to their advertisers (personalized
commercials). This major issue calls for extended NDVR services.

In brief, time-shifted TV is among the potential \emph{killer apps} of
the connected TV, but delivering time-shifted TV at large-scale under
the control of TV broadcasters is a challenge.


To address the limitations of current content delivery systems, we
explore the potentials of decentralized architectures. We distinguish
three types of elements in the system: server, clients and nodes.  A
\emph{client} is a end-user having the desire to watch the stream in a
time-shifted manner on her favorite connected device (TV, tablet,
smartphone). A \emph{server} is a front-end equipment, which is
managed by the TV service provider. Nowadays TV providers include not
only big TV broadcaster incumbents, but also online video sharing
platforms that leverages on user-generated content (for example
justin.tv and ustream) and individuals broadcasting their own
channels. A \emph{node} is an equipment under the control of the
service operator, with small storage and networking capacities. It can
be a proxy adequately deployed in the network, but it can also be a
set-top-box or the computer of a client running a video player
software with a free access to a limited part of the memory. Note that
a client can be associated with a node when a node is the box of
client or the computer of a client, but we distinguish both roles.

In Section~\ref{sec:related}, we describe the existing works related
with such decentralized (or peer-to-peer) time-shifted systems, and we
highlight their weaknesses. Then, we present two different approaches,
which aim at meeting the demands of two radically different types of
service operators.

\begin{itemize}
\item First scenario: the service provider needs a full control of the
  service delivery, but it wants to diminish the burden of its
  server. Our proposal is a peer-assisted architecture, which
  leverages on the nodes to deliver the time-shifted stream. The
  server incorporates a \emph{tracker} functionality, which aims at
  orchestrating the traffic such that the maximum number of requests
  from clients are re-routed to the nodes.  The ratio of overall
  traffic that is handled by nodes is the main performance
  indicator. We call this architecture \emph{PACUS} for Peer-Assisted
  Catch-Up Streaming. The main idea behind PACUS has been partially
  presented in~\cite{icc}. We give a detailed presentation of PACUS in
  Section~\ref{sec:pacus}.
\item Second scenario: the service provider has not the capacity to
  serve any video flows from its server. We are here in the case of
  individuals or small companies that cannot afford a well-provisioned
  data-centers. All client requests should be handled by the overlay
  of nodes. We propose a new overlay structure, which aims at
  guaranteeing that the whole stream (including the most unpopular
  programs) is always available and that any past program can be
  fetched.  We call this proposal \emph{turntable} because the system
  is based on a rotational responsibility exchange.  This structure
  has been sketched in~\cite{lcn}. We present in a comprehensive
  manner our turntable proposal in Section~\ref{sec:turntable}.
\end{itemize}


When came the time to evaluate our proposals, we faced the issue of
settings simulation parameters for a service that is not yet offered
to users, that is, without traces from real-world
applications. Indeed, neither PDVRs nor catch-up TV provide the same
flexible service as the one we envision. We have conceived a set of
synthetic traces, which are based on both previous measurements of
related TV services (IPTV and VoD services) and recent measurements of
PDVR time-shifted systems. These traces are available online for
extensive scientific
usage.\footnote{\url{http://enstb.org/~gsimon/resources/time-shifted}}
We used these traces to evaluate both proposals, as described in
Section~\ref{sec:simu}. In particular, we highlight the limitations of
the main previous work in this area (the P2TSS
system~\cite{P2TSS_Multimedia08}) and we demonstrate the benefits one
can expect from decentralized architectures for time-shifted TV
systems.



\section{Limitations of Related Works} \label{sec:related}

\subsection{Time-Shifted TV is not Video on Demand}
\label{sec:time-shifted-not}

Time-shifted streaming systems share similarities with Video on Demand
(VoD) services. An abundant literature has been published on
large-scale distributed architectures for VoD, including peer-assisted
architectures~\cite{nada,ic3n}. We raise in the following the reasons
that prevent the designers of time-shifted TV systems to implement
existing decentralized VoD systems.
 
We have already highlighted the inability of VoD servers to both
ingest and deliver a stream. Current VoD systems alternate times for
ingesting new content (during off-peak) and times for serving clients
(at peak hours)~\cite{netstitcher}. Note that Catch-up TV services,
where every program is proposed separately after it has been fully
broadcast and recorded, are a form of VoD service. Time-shifted
services however allow a end-user to time-shift a program still on air
(typically \textit{via} the popular pausing feature of PDVR). Moreover
studies have shown that most requests are for the ongoing TV
program~\cite{nielsen}. So solutions that do not consider simultaneous
ingestion and delivery are disqualified.

The length of a TV stream is several orders of magnitude longer than a
typical movie in VoD. While a movie can be considered as one unique
object, the stream of a time-shifted video is a series of
\emph{portions}, which are not uniformly popular. Some previous works
have addressed the issue of non-uniform chunk popularity in VoD
systems~\cite{5284545} but authors assume a static distribution of
chunk popularity where the first chunks are more popular than the last
ones. On the contrary, the popularity of video portions in a
time-shifted streaming system is complex because it depends on many
parameters including program popularity and broadcasting time.
Moreover, this popularity of a given portion varies with time. Recent
studies~\cite{nielsen} have shown that the average time-shift lag is
constant, so the popularity of a portion tends to decrease. Dedicated
algorithms should address this issue.

Another key difference is the volatility of clients.
In~\cite{timeshift_Globecom06}, a peak has been identified at the
beginning of each program, where many clients start streaming the
content, while the spikes of departure occur at the end of the
program. More than half of the population quits during the first ten
minutes of a program in average, and goes to another position in the
history~\cite{PPliveIPTV_TOM07}. In a same session, a shifter is
interested in several distinct portions, which can be far from each
other in the stream history. Although papers have recently addressed
the usage of pause, fast forward and rewind commands in peer-to-peer
VoD systems~\cite{Kangaroo,segmentation-aided}, no previous work
assume that it is a massive behavior of users.

\subsection{Related works}
\label{sec:related-works}

Some previous works have highlighted the problems met by time-shifted
systems based on a centralized
architecture~\cite{timeshift_ToCE08,timeshift_ToCE07,timeshift_CIT07}.
New server implementations are described in~\cite{timeshift_CIT07}.
Cache replication and placement schemes are extensively studied by the
authors of~\cite{timeshift_ToCE08}. When several clients share the
same optical Internet access, a patching technique described
in~\cite{timeshift_ToCE07} is used to handle several concurrent
requests, so that the server requirement is reduced. These works
however only partially tackle the scalability issue and they rely on a
costly infrastructure that does not meet the requirements of new forms
of small TV providers. Should the time-shifted TV service become a
common usage of connected TVs, these solutions can only satisfy a few
dominant TV broadcasters in partnerships with network operators.

Peer-to-peer time-shifted streaming systems have been presented as an
alternative architecture. In both~\cite{timeshift_P2P08}
and~\cite{timeshift_CCNC09}, a peer is a client, and every client
stores all video portions it has downloaded. This approach, called
\emph{cache-and-relay}, is also used in some peer-to-peer VoD
systems~\cite{InstantLeap}, but it does not guarantee that early or
unpopular parts are stored in the systems. A Distributed Hash Table
(DHT) allows any peer to identify the other peers that store a
requested video portion. This architecture requires that all peers
update the DHT for every portion they store. We are not convinced by
DHT-based approaches for two reasons. First, DHT are based on a
randomized hash function, which makes information about two
consecutive portions be located far from each other in the structure.
A structure that takes into account the stream linearity would be more
appropriate. Second, a peer departure conduct to multiple deletions in
the DHT: for every stored video portion that is marked in the DHT, the
peer should notify its departure.

\subsection{A focus on P2TSS system}
\label{sec:focus-p2tss-system}

The P2TSS system~\cite{P2TSS_Multimedia08} is the closest work to
ours. A backup server stores the whole video history, and a set of
\emph{peers}, both clients and servers, collaborates in order to
deliver time-shifted video portions. Each peer has a so-called
\emph{shared buffer}, where it stores the data that can be served to
other peers. A peer continuously emits two requests in parallel: one
for the video portion that it wants to play, and one for an active
caching strategy. The authors proposed two strategies for the shared
buffer management. In the \emph{initial playback caching} algorithm, a
peer that joins the system chooses a random position in the history
and downloads a full portion of the video stream from this position
until its shared buffer is full. In the \emph{live stream position
  caching} algorithm, the new peer download the current live stream
regardless of the playing position of the peer.



Peers uses a so-called pseudo-DHT, where the registered key is a
hashed index of a video position, and the value refers to the network
address of a peer storing this video position. Authors suggested that
peers only register the position of the first video chunk in the
stored portion; they however did not detail
request management. It is in particular difficult to figure out how a
peer can discover a peer storing a video portion if no peer started
downloading at this position.

\section{Preamble}
\label{sec:model-1}

\subsection{Notations and definitions}
\label{sec:vocable-notations}

We use the term \emph{piece} to refer as the fixed unit of video
stream that is used in messages, algorithms and user interfaces. A
piece is not a video \emph{chunk} because video chunks (typically
around one second video) have a too small granularity in the context
of time-shifted TV systems where stream length is in months. On the
contrary, too long pieces (hour of video) make an exchange of piece
between two nodes a long and costly process. A piece length in the
order of the minute seems reasonable, and we adopt this equivalency in
this paper. Please note that our models hereafter do not depend on
this setting.

We consider for simplicity only one channel. We denote by $\mathcal C$
the set of pieces produced by the source. Every piece is associated
with an index: the $i$th piece that has been produced by the source is
noted $c_i$. So the piece $c_0$ is the oldest piece.

The overall set of nodes that contribute to the stream delivery is
noted $\mathcal V$. At a given time $t$, only a subset of all nodes
$V_t \subseteq \mathcal V$ are active. According to the type of nodes
and the considered scenario, the variability of this subset can
change. Every node $x$ active at time $t$ stores a subset of pieces
$C_t(x)$. As long as $x$ stays in the system, it is able to serve all
pieces that it stored, with respect to its upload capacity. A node is
able to serve no more than $b(x)$ clients simultaneously. The overall
set of clients is noted $\mathcal U$. At a given time $t$, only a
subset of all clients $U_t \subseteq \mathcal U$ are active. A client
is requesting at a given time only one piece in $\mathcal C$.  The
piece requested by client $y$ at time $t$ is noted $r_t(y)$.

We call \emph{portion} a subset of consecutive pieces. When necessary,
we will denote by $P_{ij}$ the portion from $c_i$ to $c_j$, formally
$P_{ij} = \{c_k \in \mathcal C: i \leq k < j\}$, but in general we can
omit this dual subscript. The size of a portion ranges from 1 piece to
$|\mathcal C|$. Let us give an example with the P2TSS system. Nodes
store only one portion in their shared buffer. In the initial playback
caching strategy, a node $x$ randomly selects a positive integer $i$
lesser than the history time of the stream, hence, at time $t$,
$C_t(x)$ is equal to $P_{i(i+\Delta)}$ where $\Delta$ is the minimum
between the time elapsed from the arrival of $x$, and the size of the
shared buffer. In the live stream position caching strategy, $C_t(x)$
is $P_{j(j+\Delta)}$ where $j$ corresponds to the index of the piece
that was generated at the time $x$ joined the system.

A stream is said \emph{fully replicated} at time $t$ when every piece
is stored in at least one node buffer: $\mathcal C = \bigcup_{x \in
  V_t} C_t(x)$. Of course, the fact that the stream is fully
replicated does not mean that the stream is \emph{fully
  available}. For instance, the number of requests for a given piece
can be greater than the overall capacity of the nodes that store this
piece. Moreover, a piece can be stored by a node, which has no more
upload capacity because it simultaneously serves other stored pieces
to other clients. We address now the problem of managing the requests
of clients, \textit{i.e.}  assigning a node to every request, so that
the maximum number of requests can be treated.

\subsection{Use cases and dynamics of shifters}
\label{sec:use-cases-dynamics}

We detail the main events that are possible for a shifter
$x$ in a time-shifted TV system. They are usually referred to as
\emph{VCR operations}.

\emph{Pause:} it occurs when a shifter stops playing the video for a
moment, but is expected to resume streaming later from this current
position. If the client $x$ performs a pause at time $t_0$ and continues
playback at time $t_1$, the lag between $x$ and the source will
increase by $t_1-t_0$. This operation is frequently implemented in
current live streaming systems and PDVRs. In these systems, $x$
continues to download the fresh content and buffers it, which does not
allow the TV broadcaster to control the stream any longer.

\emph{Forward and Backward:} a client can perform forward or backward
in a program, or between different programs.  We distinguish these two
scenarios because both start and end times of a program are special
points where the behavior of clients have been demonstrated to be
different from other stream points.

\emph{Churn:} a client $x$ may join the system as a live client then
pause, but it can also immediately start at a past position. It is
also assumed that it can leave at any time, sometimes abruptly. We
highlight however that client leaves more frequently at the end of a
program, as shown in studies~\cite{PPliveIPTV_TOM07}.






\section{Peer-Assisted Architecture: PACUS} \label{sec:pacus}
We describe now PACUS, a peer-assisted time-shifted TV system based on
the \emph{cache-and-relay} approach. A node is associated with a
client and stores only the pieces downloaded by its client. In the
most probable scenario, a node is implemented at each client through
either the set-top-box or an installed plug-in (\textit{e.g.} the
Akamai NetSession client-side technology or the Adobe Flash Cirrus
software). A node can also be a router with caching capabilities
located near the clients, as it has been recently proposed in
Content-Centric Networks~\cite{li11icc}. The server has two
functionalities: a backup server in case no node can fulfill a
request, and a \emph{tracker}, which is in charge of forwarding the
clients' requests to appropriate nodes. Note that the tracker can
redirect requests to its own servers when necessary, for instance to
adapt commercials to the actual watching time. The main mission of the
tracker is to provide to every client $y$ a set of nodes having the
piece requested by $y$.

\subsection{Background}
\label{sec:background}

The problem of maximizing the number of fulfilled requests at a given
time is formulated as follows. Let $G=(U \cup V,L)$ be a bipartite
graph where $U$ and $V$ are the clients and the nodes respectively,
and the set of edges $L \subseteq U \times V$ contains an edge between
$y \in U$ and $x \in V$ if and only if the node $x$ stores the piece
requested by $y$, formally $yx\in L \Leftrightarrow r_t(y) \in
C_t(x)$. The goal is to assign nodes to clients with the constraint
that no node $x$ can be used more than $b(x)$ times, so that the
maximum number of clients is served.

We abusively note $L(\cdot x)$ the set of clients the node $x$ can
serve (\textit{i.e.} $x$ has the piece requested by every client in
$L(\cdot x)$), and similarly $L(y \cdot)$ for the set of nodes that
can serve the client $y$ (\textit{i.e.} the nodes in $L(y \cdot)$ have
the piece requested by $y$). Let define the vector $l$ in
$\{0,1\}^{|L|}$ such that:
\begin{equation*}
l_{yx} = \left\{
\begin{array}{ll}
1 & \textrm{ if } x \textrm{ is assigned to } y,\\
0 & \textrm{ otherwise}.
\end{array}
\right.
\end{equation*}

The optimal assignment from nodes to clients is a solution to the
following system of linear inequalities
\begin{align}
  \label{eq:sp:matching:left} & \sum_{x \in L(y \cdot) } l_{yx} \leq 1 && \textrm{ for } y \in U,\\
  \label{eq:sp:matching:right} & \sum_{y \in L(\cdot x)} l_{yx} \leq b(x) && \textrm{ for } x \in V,\\
  \label{eq:sp:matching:non-negativity} & l_{yx} \geq 0 && \textrm{
    for } yx \in L.
\end{align}

This problem is in the family of semi-perfect matching
problems. Efficient polynomial-time algorithms ensuring a fair
allocation to the nodes can be found in~\cite{matching06}. This
problem has also been studied in its online case when requests arrive
iteratively~\cite{kalyanasundaram2000optimal}. An optimal assignment
at a given time can thus be computed in a reasonable time. This
computation requires however a complete view of the system, (requests
from clients, node capacity and storage buffer). When the service
architecture contains a front-end server controlling well-configured
servers, this assumption is reasonable.

\subsection{Leveraging on a multiple-interval modeling}

We look for faster algorithms that do not require recomputing
the whole assignment after every piece exchange. We model the system
as a \emph{multiple-interval graph}. We first recall the main
principle behind this structure, and then we explain why this
structure is adequate for time-shifted streaming systems.

A graph is called an \emph{interval graph} if its vertices can be put
in a one-to-one correspondence with a family of intervals on the real
line, such that two vertices are adjacent if and only if their
corresponding intervals have nonempty intersection.  Formally, let
$\{I_1, I_2, \cdots ,I_n\}$ be a set of intervals.  A graph $G_i =
(V,E)$ is an interval graph if, for any pair $u, v$ of vertices, we
have $I_u \cap I_v \neq \emptyset \Leftrightarrow uv \in E$.  A
natural generalization of interval graphs is the multiple-interval
graph.  A multiple-interval graph is an intersection graph of a family
of several intervals.  Formally, let $I(u)$ be the set of intervals
$\{I_{1u}, I_{2u}, \cdots ,I_{ku}\}$ and $I(v)$ be another set of
intervals $\{I_{1v}, I_{2v}, \cdots ,I_{k'v}\}$. If there are two
intervals $I_{iu} \in I(u)$ and $I_{jv} \in I(v)$ such that $I_{iu}
\cap I_{jv} \neq \emptyset$, we say that $I(u)$ and $I(v)$
\emph{intersect}.  A graph $G_I = (V,E)$ is a multiple-interval graph
if, for any pair $u, v \in V$ of vertices, we have: $ uv \in E
\Leftrightarrow I(u) \textrm{ and }I(v) \textrm{ intersect } $.
Problems on multiple-interval problems find typical applications for
multi-task scheduled problems or resource allocation problems. This
structure has been extensively studied. Recent results can be found
in~\cite{multiple_interval}.

The equivalence between an interval (in multiple-interval graph) and a
stream portion (as we defined it previously) is immediate.  Because of
VCR operations, each node store multiple portions, so each node is
associated to multiple intervals (note that the P2TSS system can be
modeled as a simple interval-graph because each peer only has one
interval, its shared buffer). By storing information about video
portions and nodes in a multiple-interval structure, a tracker can use
any of the existing fast deterministic
algorithms~\cite{t_interval_exactOptimal} to determine all nodes
having a given piece in a time that is linear of the number of stored
portions.

\subsection{Communications}

\paragraph*{Nodes to tracker communications}
Using the same technique than P2TSS, the tracker in PACUS records the
two endpoints of each video portion stored by the nodes. Thus, every
client should notify the tracker only for the time-shifting event
(\textit{i.e.}, join, pause, fast-forward, rewind and leave). Because
the playback rate is constant, the tracker can infer at any time the
index of the piece that is currently played by the client.  When a
client ends playing the video portion (it leaves or switches again),
the tracker records the last requested piece, then, it adds the
interval (\textit{a.k.a.}  video portion) in the data structure.




\paragraph*{Tracker to Client Communication}

The mission of the tracker is to provide to every client issuing a
request for a piece a set of nodes having a high probability to be
able to upload this piece, not only because they store this piece, but
also because their uplink is not congested. The optimal assignment,
which requires a re-computation after every piece, being too costly to
implement in the context of a large-scale service, we propose for
PACUS a lightweight solution where the tracker sends a subset of all
nodes that are able to serve a client for a piece, then the tracker
refreshes this subset of nodes only when a time-shifting event is
notified by the client. The multiple-interval structure allows to
determine all nodes storing the piece. From this set of nodes, several
strategies can then be implemented in order to select a subset of
nodes:
\begin{itemize}
\item \emph{Random}: the nodes are randomly selected
\item \emph{Network}: a tracker that knows the location of nodes and
  clients in the network can determine the subset of nodes that
  minimize the distance (in term of latency or number of traversed
  Autonomous Systems) from the nodes to the client.
\item \emph{Available Capacity}: a tracker being notified by the nodes
  about their available capacity to serve a client can determine the
  subset of nodes that have the more available capacity.
\item \emph{Playback}: the tracker can choose the nodes storing the
  longest video portion from the requested piece. This strategy, which
  does not require any additional knowledge, may ensure that the next
  pieces requested by the client can be served by the same subset of
  nodes.
\end{itemize}

\paragraph*{Peer to Peer Communication}

When a client receives from the tracker a list of expectedly active
nodes, it can decide to download the requested piece from one or
several of these nodes. Many recent works have dealt with efficient
stream delivery from a set of servers. We prefer here to let the
system designer decide the best algorithm, which depends on many
parameters, typically the video encoding. In our simulation, we
implemented the simplest solution: the client chooses randomly one
node, which is in charge of delivering the piece if it has the
capacity to do it. Then, the client keeps on downloading from this
node the video portion until this node is no more able to deliver new
pieces (its stored video portion terminates or it leaves). The client
can then contact the other nodes in the set of nodes, if any.


\section{Fully-Distributed Architecture: Turntable}\label{sec:turntable}

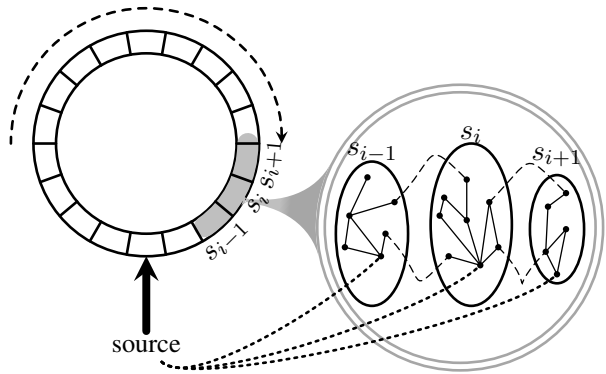
\begin{figure}[!t]
\centering
\usetikzlibrary[shapes]
\usetikzlibrary{mindmap} 
\usetikzlibrary[mindmap]
\begin{tikzpicture}[scale=0.6,cap=round,line width=1pt,>=stealth]

\node (n1) at (1.7,-1.2) [circle,minimum size=0.5cm,draw=red,fill,thick,white] {};
\node (n22) at (-15:7.2) [circle,minimum size=3.8cm,draw=gray!70] {};
\node (n42) at (-15:7.2) [circle,minimum size=3.6cm,draw=gray!70] {};
\node (n3) at (0,0) [circle,minimum size=3cm,draw=black] {};
\path (n1) to[circle connection bar switch color=from (gray!50) to (gray!70)] (n22);

\draw (0,0) circle (2cm);
\draw [<-, line width=1pt, dashed] (0,0) (0:3cm) arc (0:180:3cm);
\draw [gray!50, line width=2.8mm] (0,-2) (0:2.25cm) arc (0:-55:2.25cm);

\foreach \angle / \label in
  {0/1, 20/2, 40/3, 60/13, 80/12, 100/11, 120/10, 140/9,
   160/8, 180/7, 200/6, 220/5, 240/4, 260/14, 280/15, 300/16, 320/17, 340/18}
{
  \draw[line width=1pt] (\angle:2.5cm) -- (\angle:2cm);
}

\draw [] (1.3,-1.7) node[rotate=45,below=4pt] (si) {$s_{i-1}$};
\draw [] (1.9,-1.0) node[rotate=60,below=4pt] (si1) {$s_{i}$};
\draw [] (2.1,-0.4) node[rotate=85,below=4pt] (si2) {$s_{i+1}$};

\draw [] (5,-0.2) node[] () {$s_{i-1}$};
\draw [] (7.2,0.2) node[] () {$s_i$};
\draw [] (9.1,-0.3) node[] () {$s_{i+1}$};

\node[inner sep=3pt] at (0,-4.5) (source) {\small source};
\draw[line width=3pt, -stealth] (source) -- (0, -2.5);
\tikzstyle{every node}=[fill=gray!100,circle,inner sep=3pt]
\tikzstyle{level 1}=[level distance=6mm]
\tikzstyle{level 2}=[level distance=5mm]
\tikzstyle{level 1}=[sibling distance=15mm, set style={{every node}+=[]}]
\tikzstyle{level 2}=[sibling distance=6mm, set style={{every node}+=[]}]
\tikzstyle{level 3}=[sibling distance=2mm, set style={{every node}+=[fill=blue]}]

\draw (5,-2) ellipse (8mm and 16mm);
\draw (7.2,-1.8) ellipse (9mm and 18mm);
\draw (9.1,-1.9) ellipse (6mm and 12mm);

\fill [] (7.4,-2.7) circle (2pt);
\fill [] (6.7,-2.5) circle (2pt);
\fill [] (7.6,-1.3) circle (2pt);
\fill [] (7.8,-2.3) circle (2pt);
\fill [] (7.1,-1.7) circle (2pt);
\fill [] (6.6,-1.2) circle (2pt);
\fill [] (6.5,-1.6) circle (2pt);
\fill [] (7.1,-0.8) circle (2pt);
\draw [line width=0.5pt] (7.4,-2.7) -- (6.7,-2.5)
			 (7.4,-2.7) -- (6.5,-1.6) --(6.6,-1.2)
 			 (7.4,-2.7) -- (7.1,-1.7) -- (7.1,-0.8)
			 (7.4,-2.7) -- (7.8,-2.3) -- (7.6,-1.3)
			 (7.1,-1.7) --(6.6,-1.2)
			 (7.4,-2.7) -- (7.6,-1.3);

\fill [] (4.9,-0.75) circle (2pt);
\fill [] (5.5,-1.3) circle (2pt);
\fill [] (4.5,-1.6) circle (2pt);
\fill [] (5.3,-2) circle (2pt);
\fill [] (4.4,-2.3) circle (2pt);
\fill [] (5.2,-2.5) circle (2pt);
\draw [line width=0.5pt] (5.2,-2.5) -- (4.4,-2.3) -- (4.5,-1.6)
			 (5.2,-2.5) -- (5.3,-2)
			 (5.2,-2.5) -- (4.5,-1.6)
			 (4.5,-1.6) -- (5.5,-1.3)
			 (4.5,-1.6) -- (4.9,-0.75);

\fill [] (9.3,-1.1) circle (2pt);
\fill [] (8.9,-1.4) circle (2pt);
\fill [] (9.3,-1.9) circle (2pt);
\fill [] (8.85,-2.4) circle (2pt);
\fill [] (9.1,-2.9) circle (2pt);
\draw [line width=0.5pt] (9.1,-2.9) -- (8.85,-2.4) -- (8.9,-1.4)
			 (9.1,-2.9) -- (9.3,-1.9) --(8.85,-2.4)
			 (8.9,-1.4) -- (9.3,-1.1);

\draw [dotted, line width=1pt] (source) ..controls+(1,-1) and +(-1,-1) ..(5.2,-2.5);
\draw [dotted, line width=1pt] (source) ..controls+(1,-1) and +(-1,-1) ..(7.4,-2.7);
\draw [dotted, line width=1pt] (source) ..controls+(1,-1) and +(-1,-1) ..(9.1,-2.9);

\draw [densely dashed,line width=0.5pt] (5.5,-1.3) ..controls+(1,1) and +(-1,1) ..(7.1,-0.8);
\draw [densely dashed,line width=0.5pt] (5.3,-2) ..controls+(1,-1) and +(-1,-1) ..(6.7,-2.5);

\draw [densely dashed,line width=0.5pt] (7.6,-1.3) ..controls+(1,1) and +(-1,1) ..(9.3,-1.1);
\draw [densely dashed,line width=0.5pt] (7.8,-2.3) ..controls+(1,-1) and +(-1,-1) ..(8.85,-2.4);
\end{tikzpicture}
\caption{The distributed turntable structure}\label{fig:turntable}
\end{figure}

We describe now the turntable, an overlay of nodes that ensure a full
replication of pieces in a fully distributed manner. Our turntable
structure is represented in Figure~\ref{fig:turntable}. The basic idea
is as follows. We partition the set of nodes into several subsets,
every subset being responsible of the same set of video
portions. These subsets of nodes form what we call \emph{intra-sector
  overlays}. Then, we connect these overlays through
\emph{inter-sector links} so that the whole structure is connected.

Among the advantages of the turntable structure, we highlight that
every sector participate at the same level, so the nodes contribute
almost evenly to the delivery of the pieces. Because every sector is
associated with a video portion of fixed size, the piece popularity
variation does not impact the sector request load. Moreover, every
sector should store the same amount of pieces. Besides, the
maintaining of the structure is not heavy as the only fundamental
principle is to have a contact with a node in the next sector. We
emphasize that various algorithms can be used. Contrarily to previous
works, the design of this structure fits the characteristics of
time-shifted streaming systems.




The turntable is divided into $m$ \emph{sectors} noted $s_i, 0\leq i <
m$.  Every node joins exactly one sector. The turntable implements a
rotational motion in clockwise direction. Every sector is responsible
of some video portions of size $\Lambda$ pieces. We call \emph{cycle}
the time corresponding to $\Lambda$ pieces. At every cycle $t$, the
source produces a new video portion ($\Lambda$ pieces), which is sent
to a sector, say $s_i$, then the portion produced at cycle $t+1$ is
sent to the sector $s_j$ with $j=(i+1)\mod m$, and so on\footnote{in
  the following, we omit the modulo for notation clarity.}.  Hence,
every video portion is under the responsibility of a sector,
\textit{i.e.}, of a subset of nodes. They are responsible to store all
pieces of every portion given by the source, and to deliver them to
clients. 

A client is connected to a set of nodes, which are expected to be able
to serve it. These nodes belong to the sector corresponding to the
piece that the client is willing to download.  From a cycle $t$ to a
cycle $t+1$, this set is refreshed, because, unless a client has
paused the stream, the pieces requested during cycle $t$ are different
from the pieces requested at cycle $t+1$. When a client wants to
download any past portion of the stream, it should first determine the
sector associated with the first piece of this portion.  After it
finds a node that has stored the requested pieces in this sector, it
can start the downloading, then, it should jump to the next sector in
order to retrieve the next pieces and continues consuming the stream.

The source must be connected to at least one node in every sector.
These nodes are called \emph{representative}.  We denote by
$\hat{x}_i$ the representative for sector $s_i$.  When it is time for
a sector $s_i$ to handle a new video portion, the source alerts the
representative $\hat{x}_i$ and sends the first piece of this portion
to it, then this piece, as well as the $\Lambda$ following ones, are
diffused in the sector. During the $m$ next cycles, this video portion
is called the \emph{fresh portion} for this sector.

\subsection{Overlay Construction}

The structure of the intra-sector overlay is not constrained. The
primary goals of intra-sector overlay are to fairly distributed the
pieces among the nodes, to allow fast discovery of nodes that store a
given piece, and to ensure the full replication of all video portions
under its responsibility. A huge set of works related to peer-to-peer
data sharing has addressed such issues for a decade. For our
implementation, we used a gossip-based technique inspired by
T-Man~\cite{t-man}.  Every node was connected to a subset of nodes,
which it continuously refreshed.  Nodes exchanged messages on a
periodic manner, these messages carrying neighborhood
information. Then, every node connected to the ``best'' nodes among
its current neighbors and all the possible neighbors that were
described in these messages. The resulting overlay structure let every
node be quickly connected to the nodes that it considers as the
best. For us, the best nodes are the closest ones in the network.

The goal of inter-sector links is to ease the retrieval of consecutive
pieces for clients.  A client $y$ retrieving a stream portion that is
longer than one cycle produces successive requests to consecutive
pieces in consecutive sectors.  The purpose of an inter-sector link is
to connect two nodes that are in consecutive sectors and that store
some successive past video pieces. Consider a client $y$ that
downloads the last piece of the video portion stored in a sector by a
node $x_i \in s_i$. The next request of $y$ will be in sector
$s_{i+1}$, so $y$ should find in sector $s_{i+1}$ a node that stores
the piece next to the one it just downloads from $x_i$.  Ideally, the
next piece is stored by the node $x_{i+1}$ which is an inter-sector
neighbor of $x_i$ in sector $s_{i+1}$. In this case, $x_i$ can
directly introduce $x_{i+1}$ to $y$.  In our simulations, we aimed at
both maximizing the configuration where nodes from consecutive sectors
own consecutive pieces, and again minimizing the network cost.  First,
if the node $x$ had a piece $c_i$, it preferred a node $x'$ in the
next sector that stored the piece $c_{i+1}$. Then, among all nodes it
knew, $x_i$ selected the closest nodes in the network.

\subsection{Algorithms} \label{subsec:algo}

We describe now the algorithms that are implemented on top of the
turntable overlay. Please note that various protocols can be
designed. We present here the ones that have demonstrate good
performances during our simulations. First, an algorithm for the
\emph{diffusion of fresh portions}. As the fresh pieces are also the
most requested pieces, they should be diffused as quickly as possible
to many nodes within the sector. Second, an algorithm ensuring \emph{a
  fair repartition of the past pieces}. Ideally, the number of
replicas of a piece should correspond to the number of requests
emitted for this piece. The other algorithms, for example finding a
node that stores a requested piece, have to be designed with respect
to the choice of the intra-sector overlay structure. We focus here on
the algorithms that are specific to the turntable time-shifted TV system.

\paragraph*{Fresh piece management}


Initially, a representative $\hat x$ of a sector receives the first
piece of a fresh portion from the source. The representative delivers
this piece into a subset of its intra-sector neighbors. We then
implement a gossip process based on two parameters: a portion can be
forwarded up to a certain number of times, and each forwarding node
sends it to a subset of its intra-sector neighbors according to a
forwarding probability.  The computation of this probability parameter
is an issue because some fresh portions can be popular because they
are the initial portions of a program, or because the system is at a
peak of the number of clients. In this case, a lot of replicas of the
fresh portion should be produced as quickly as possible. On the
contrary, portions at the end of a program or during period with few
users should not necessarily be heavily replicated.

Our approach consists in leveraging on the observed popularity of
fresh pieces in the previous sector to adjust the probability
parameter. We denote by $f_j$ the failure frequency of a fresh piece
$c_j$, \textit{i.e.} the number of requests for $c_j$ that have not
been fulfilled. The idea is that $f_j$ is with high probability
similar to $f_{j-1}$ if the forwarding probability to diffuse the
piece $c_j$ is the same as the one for $c_{j-1}$. We first have to
calculate a failure frequency of a piece in a sector. This estimation
depends on the algorithm that is implemented to issue a request for a
piece in a sector. In our implementation, we used a $k$-neighborhood
flooding to request pieces in an intra-sector overlay, hence every
node was able to estimate failure frequency of every piece by
trivially counting the number of received requests for this piece. In
order to adjust the forwarding probability, we used an
\textit{Additive Increase Additive Decrease} mechanism.  When a node
$x$ was notified of a ratio of request failures lower than a given
threshold $f_{low}$, it decided to decrease the probability. On the
contrary, a ratio of request failures higher than a given threshold
$f_{high}$ resulted in a probability increase.

Please note that the case that the probability is already equal to
one, but the failure frequency is above the threshold is a good
indicator that the sector is under-provisioned, so new nodes should
preferentially be allocated to this sector.

\paragraph*{Past piece management}

The limitation of the storage capacities imposes to not create a
replica of each piece at each node.  An important issue for past piece
management is the choice of the video piece to be removed when the
local storage is full and when a new video piece should be stored.
Classic caching policies, in particular \emph{least recently used}
(LRU) and \emph{least frequently used} (LFU), can be used to replace
old pieces. In our implementation, we chose the LRU policy because
this policy has exhibited good performances in peer-assisted VoD
systems~\cite{simpleCache}. However a key objective is to guarantee at
least one replica for every piece. Therefore the algorithm that we
actually implemented is a \emph{pseudo-LRU} algorithm where a node
first established a list of video pieces that it can discard because
it was sure that at least one other replica existed in its sector,
then, it determined the portion to remove among this selected portions
by the LRU policy.

In order to guarantee the full replication when nodes can leave, we
implemented an algorithm that is inspired by~\cite{P2R2_PODS2008}
where authors describe a distributed replication algorithm with regard
to the popularity of data item and storage capacity of nodes, as well
as the heterogeneity and dynamics of network and workload. Every node
estimated the past video pieces that required new replica, and, when
this node was not fully occupied by serving clients, it requested one
of these pieces. Such a piece management can typically be performed
during off-peak.



\section{Simulations} \label{sec:simu}
We implemented both proposals on the PeerSim simulator~\cite{peersim}
to evaluate their feasibility and performances. We used a set of
synthetic traces that we first describe. Then we present some general 

\subsection{Synthetic traces and general simulation settings}
\label{sec:synthetic-traces}

We utilized two sets of studies conducted in 2008 and 2009. The first
set is real measurements by PDVR vendors. A major actor, namely TiVo,
gives regularly data about the usage of its
set-top-boxes.\footnote{http://stopwatch.tivo.com} The paper that we
used in priority is a Nielsen report~\cite{nielsen}, which gives
precious insights about user behavior. The second set of related works
is the measurements conducted on IPTV~\cite{IPTVMeasure_sigcom08} and
VoD systems~\cite{VoD_sigops06}.


\begin{itemize}
\item \emph{Program Popularity}: it has been established that the
  program popularity is mostly a function of its broadcast
  hour. In~\cite{nielsen}, a quarter of shifters have a stream lag
  that is less than one hour, around $40\%$ of them watch their
  program less than 3 hours after the live program, and more than half
  of shifters are enjoying a program that has been broadcasted the
  same day. Of course, some programs can be more popular than others,
  but this popularity is a consequence of the time at which they are
  broadcasted and of the number of shifters that are active at that
  time. Therefore, we pre-fixed a popularity parameter to the programs
  but we limited the impact of this parameter on the shifter behavior.

\item \emph{Churns and Switches}: in~\cite{timeshift_Globecom06}, a
  peak has been identified at the beginning of each program, where
  many viewers start streaming the content. Then, similarly as in VoD,
  the spikes of departure occur either at the end
  of the program, or because the user does not find any interest after
  browsing the beginning of the
  program~\cite{PPliveIPTV_TOM07}. Moreover, in most cases, the more
  popular is the program, the shorter is the session length.
  According to these facts, we decided to assign a role to every
  client: half of them are \emph{surfers} (watch a same program during
  $1$ or $2$ minutes before to switch to another program), $40\%$ of
  them are \emph{viewers} (switch after a duration uniformly chosen
  between $2$ and $60$ minutes), and only $10\%$ are \emph{leavers}
  (stay on a program during a time comprised between $60$ and $1800$
  minutes, \textit{i.e.} a TV constantly opens during up to $30$
  hours).
\item \emph{Time-Shifted Usage by Hour}: the TV 
  prime-time is on evening. Measurements made
  in~\cite{nielsen} confirm that shifters are more connected
  at certain time of the day than others. In US, only $1\%$ of
  shifters start shifting between 6:00~AM and 7:00~AM. On the
  contrary, more than $11\%$ join the system between 9:00~PM and
  10:00~PM (note that shifters begin watching live programs before
  time-shifting, typically after a pause or a rewind through immediate
  past content).
\end{itemize}

\begin{table}
\centering
\begin{tabular}{ccc|ccccc}
 \multicolumn{3}{c}{\emph{scenarios}} & \multicolumn{5}{|c}{\emph{node
     capacities}} \\
 \emph{content} & \emph{distrib.} & \emph{nickname} & \emph{0} & \emph{1} & \emph{2} & \emph{3} & \emph{4} \\ \hline \hline
 \multirow{2}{*}{HDTV} & homoge. & \textbf{h-HD} & 25\% & 50\% & 25\% & 0\% & 0\% \\
 & heteroge. & \textbf{H-HD} & 40\% & 30\% & 20\% & 10\% & 0\% \\ \hline
 \multirow{2}{*}{IPTV} & homoge. & \textbf{h-IP} &  0\% & 25\% & 50\% & 25\% & 0\% \\
 & heteroge. & \textbf{H-IP} &  0\% & 40\% & 30\% & 20\% & 10\%

\end{tabular}
\caption{Percentage of nodes able to deliver a given number of streams} \label{tab:capacities}
\end{table}

From these observations, we designed some synthetic traces that
simulates the behavior of clients. These traces describe a set of
clients, which alternate off and on periods, and their request for
past pieces. At peak, around $1,\!000$ clients are active, while the
number of concurrent shifters is around $200$ during off-peak.  The
other parameter for the simulation are as follows.  At every cycle, a
new piece, \textit{i.e.} the basic video unit, was generated.  A piece
represented one minute of stream (both terms piece and minute are used
alternately).  The number of cycles was $30\,000$, \textit{i.e.}, more
than $20$ days. Several continuous pieces formed a program.  Every
program was associated to a \emph{genre}, a popularity chosen in a
predefined distribution, and a length in $[30,100]$.  Three genres
were considered: $80\%$ were \emph{free}, $15\%$ were \emph{news} and
$5\%$ were \emph{kids}.  As it has been noticed in various IPTV
measurements, a viewer is more likely to choose a program in the same
genre when it switches.


In our system, one piece transmission consisted of one-minute long
stream delivery. Therefore, the capacity of a peer was described as
the number of concurrent streams that the node is able to send to
clients. If the stream source generated High-Definition TV (HDTV)
content, the capacity of peers was smaller than if the video was a
classic IPTV format. We simulated both scenarios. In the HDTV
scenario, the average capacity of nodes was $1$ (in average, a node
was able to send one stream to only one client), although they was
able to serve in average $2$ clients in the IPTV scenario. For the
distribution of upload capacities, we considered two scenarios, where
the distribution of capacities was either \emph{homogeneous} (most
nodes had the same capacities), or \emph{heterogeneous} (many nodes
had no capacity, but some powerful nodes were able to deliver a
lot). We describe in Table~\ref{tab:capacities} the four considered
scenarios with their nicknames.

Finally, we have to set the location of nodes and clients, which
depends on the type of service provider.  If the TV channel is an
international broadcaster watched from all over the world, say
\textit{e.g.} CNN, clients are almost uniformly spread over many ASes.
On the contrary, many broadcasters are essentially local, in the sense
that their programs aim at being watched by a population that is
geographically well-identified.  In this simulator, we configured an
Internet map from a CAIDA data-set containing $28,\!421$ ASes and their
relationship (\emph{peering} or \emph{transit}).  We considered the
first configuration, \textit{i.e.} a worldwide service.  For every
simulation, $1,\!000$ ASes were randomly selected among all $28,\!421$
ASes in the dataset. Nodes and clients were uniformly spread over
these ASes.

\begin{figure}[ht]
\centering
\includegraphics[width=0.6\columnwidth]{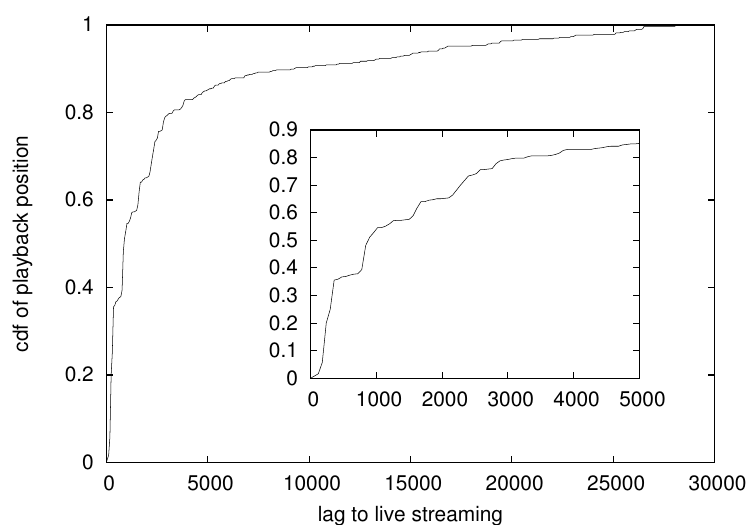}
\caption{Distribution of playing positions}\label{fig:playingposition}
\end{figure}

\paragraph*{Validation}
Figure~\ref{fig:playingposition} represents the Cumulative
distribution Function (CDF) of the lag of shifters at the end of our
simulation.  The embedded plot zooms on the $5,\!000$ first minutes,
which represents actually more than $80\%$ of shifters.  A point at
$(1,\!000,0.50)$ means that half of shifters are watching a program
broadcasted less than $1,\!000$ minutes ago. Note on the embedded
figure that variable program popularity results in a sinuous
curve. This curve is actually conform to the measurements made
in~\cite{nielsen}.

\subsection{Simulations on the PACUS system}

\begin{figure}[!t]
\centering
\includegraphics[width=0.6\columnwidth]{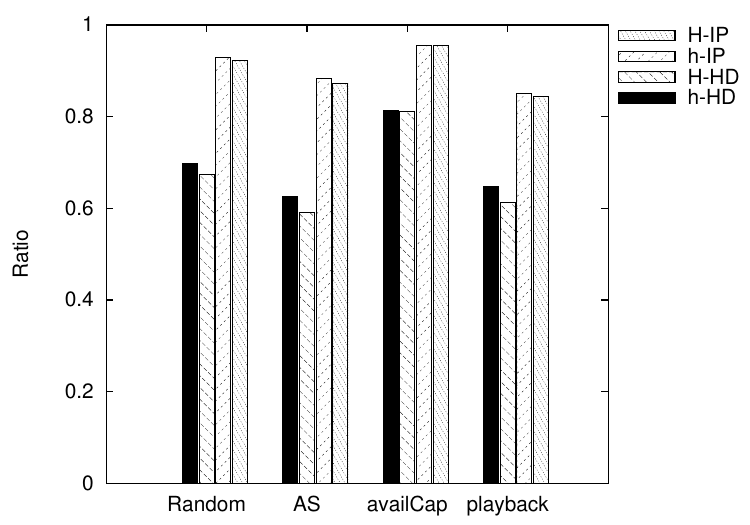}
\caption{Ratio of pieces retrieved from nodes} \label{fig:dataRetrieve}
\end{figure}

We first aim to determine the best tracker strategy for selecting the
subset of nodes.  We present in Figure~\ref{fig:dataRetrieve} the
percentage of piece requests that are handled by a node for the four
strategies (\emph{AS} stands for the strategy where network distance
prevails, \emph{playback} is for the longest video portion and
\emph{availCap} is for a selection of nodes based on their available
upload capacity).

We highlight three observations.  First, the ratio of requests handled
by the nodes is large. In the IPTV scenario, this ratio represents
more than $80\%$ of requests. More than half of requests can also be
treated in the HDTV scenario. The server can actually be used as a
backup server or for other purposes.  Second, as can be expected, the
strategy where the tracker is able to know the available capacity of
every node outperforms the other strategies. We notice the relatively
poor performances of the strategy based on the proximity of playback
position. Despite its flaws, the random strategy has better results
than the network-friendly one. We uses the random strategy in our
comparative evaluation.  Third, the heterogeneous distribution of peer
capacity does not affect the capacity of the overlay of nodes. Even
for strategies agnostic of the upload capacity, results between
homogeneous and heterogeneous distributions are slightly dropping, but
the drop is negligible.

\begin{table}
\centering
\begin{tabular}{c|cc|c}
  & \emph{PACUS} & \emph{Centralized} & \emph{Difference} \\ \hline \hline
  \textbf{HDTV scenario} & $6.12$ & $13.28$ & $-53.9\%$ \\
  \textbf{IPTV scenario} & $3.57$ & $13.28$ & $-73.1\%$
\end{tabular}
\caption{Average number of traversed AS by piece} \label{tab:as}
\end{table}

Then, we study the overall impact on the network (see
Table~\ref{tab:as}). PACUS are compared with a centralized system with
only one server. For every piece request, we count the number of ASes
traversed by the piece in PACUS, and compare it to the number of ASes
traversed by the piece when it comes from the server. Hence, we
measure the overall cross-domain traffic generated by the time-shifted
TV system. The PACUS results are obtained by the network-friendly
policy where the tracker chooses the ``closest'' nodes such that the
number of traversed ASes between them is the smallest. We obtained
noteworthy results with gains that reach up to $73\%$. Actually, a
peer-assisted architecture is not only a way to reduce the traffic at
the server side. It is also an architecture for reducing the overall
traffic over the Internet.

\subsection{Simulations on the Turntable System}

We focused now on the settings of the turntable system.  We fixed
$m=20$ sectors, each of which having one representative.  Each node,
randomly assigned to a sector, can store 100 pieces.  A node is
connected with $5$ intra-sector neighbors, and $5$ inter-sector
neighbors that are both chosen among 20 acquaintances. At every cycle, a
client should reset the set of peers that it met in the sector
of the last piece it downloaded.  Then, it looked for the next pieces in
the next sector following the inter-sector links.
In this first set of results, we focus on the piece
replication, the piece requests, and the ratio of flooding and failures.

\begin{figure*}[t]
\centering
\subfigure[Number of fresh pieces after $m$ cycles]{\label{fig:freshPieceDistr}
\includegraphics[width=0.45\columnwidth]{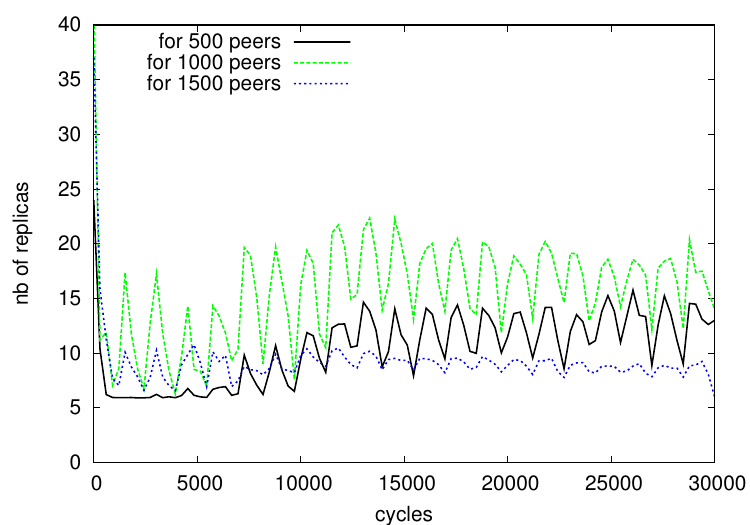}
}
\subfigure[Evolution of the number of piece replicas over time]{\label{fig:pieceTrend}
\includegraphics[width=0.45\columnwidth]{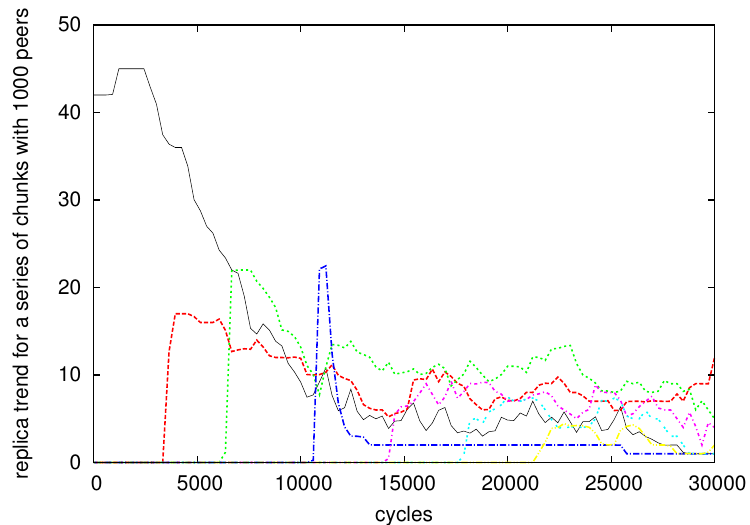}
}
\caption{Piece replication in the turntable structure}
\end{figure*}

\paragraph*{Piece Replication}

We represent in Figure~\ref{fig:freshPieceDistr} the evolution of the
number of replicas that are generated for the fresh pieces during the
$m$ cycles following its production.  As can be observed, the number
of piece replicas varies with the number of concurrent clients. When
the number of clients increases, the number of failures increases too,
so the forwarding probability tends to become higher. This result
highlights that our algorithm is able to adopt adequate reactions to
environment changes. We observe also that the variation of the
forwarding probability is less intense when the number of nodes is
$1,\!500$. The reason is that the failures are less important for this
number of peers.

Then, we show the evolution of some randomly chosen pieces in
Figure~\ref{fig:pieceTrend}. The number of replicas of the oldest
piece tends to increase quickly in comparison to other pieces. In the
earliest times, the forwarding probability is equal to 1, hence the
first pieces are almost broadcasted within one sector. Then the
forwarding probability decreases and stabilizes, so the high level of
replication after the piece production is limited.  The evolution of
pieces follows a normal evolution according to the demand. A produced
piece is popular after it is generated, so the number of replicas
increases quickly, then its popularity decreases slowly, as well as
its number of replicas.

\paragraph*{Quality of Services - Fulfilled Requests}

\begin{figure*}[!ht]
  \centering
\subfigure[Ratio of flooding per
  demand]{\label{fig:floodRatio}
\includegraphics[width=0.48\columnwidth]{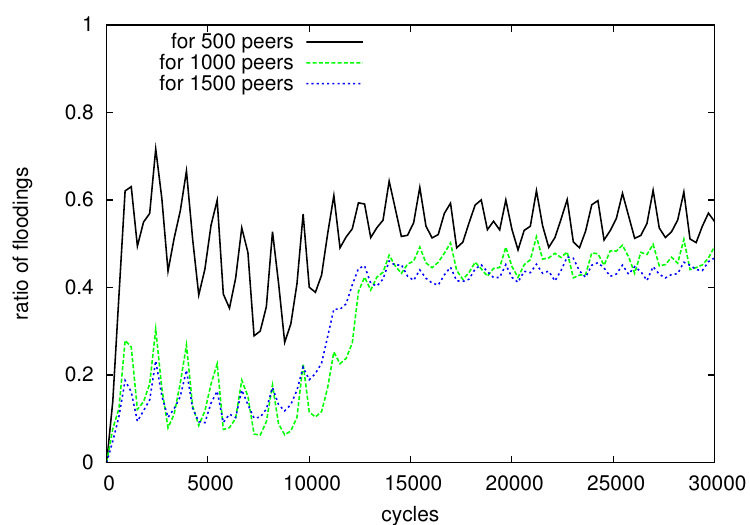}
}
\subfigure[Ratio of failures per
  demand]{\label{fig:failureRatio}
\includegraphics[width=0.48\columnwidth]{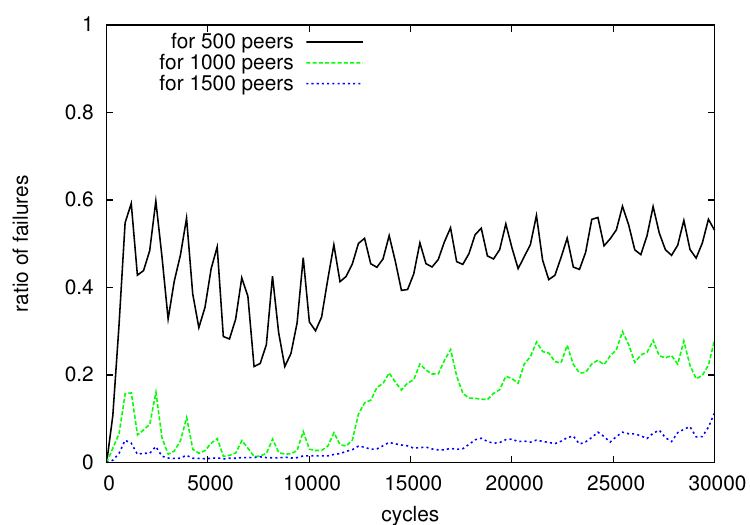}
}
\caption{Quality of services in the turntable structure}
\end{figure*}

We now observe the quality of service. We distinguish a
\emph{flooding}, when a client is connected to a set of nodes that
cannot treat its request, so the client has to issue a request (in our
implementation, a $k$-hop flooding), and a \emph{failure}, when the
client is unable to find any node able to treat its request. In the
former case, the overhead generated by the request messages is
important. In the latter case, the system is unable to serve the
client. Figures~\ref{fig:floodRatio} and~\ref{fig:failureRatio} show
the evolution of the ratio of both parameters to the number of
received requests during the simulation.


The number of nodes has a dramatic impact on the quality of
services. When the number of nodes is equal to $1,500$ (this number
has to be compared with the number of concurrent clients), the number
of failures is still low (less than $5\%$ of requests are not
fulfilled). Actually, most failures occur for very fresh pieces,
where, despite the algorithm for the diffusion of fresh pieces, nodes
have not the capacity to generate the number of replicas on time. For
smaller number of nodes, the problem of congestion becomes more
important.  With an average upload capacity of 1, nodes can only serve
as many clients as $n$. What we show is that the ratio of requests is
approximately the ratio number of clients to number of nodes. It means
that the system is able to almost entirely utilize the upload
resources of nodes.

In these figures, we also observe a sharp increasing of ratio of
flooding (and failures) after a given number of cycles. This cycle
marks the time at which it becomes physically impossible to find all
pieces in the immediate neighborhood. Indeed, nodes have only 5
neighbors, and the storage capabilities of peers is limited to 100
peers.

\subsection{Comparative studies}

We now compare three systems: our proposals PACUS and turntable and
the previous works P2TSS.  We compare the simulation results of piece
replication algorithms and load gains at the server side.

\paragraph*{Piece Replication}

\begin{figure*}[!ht]
\centering
\subfigure[PACUS]{\label{fig:pieceReplicasForPacus}
\includegraphics[width=0.48\columnwidth]{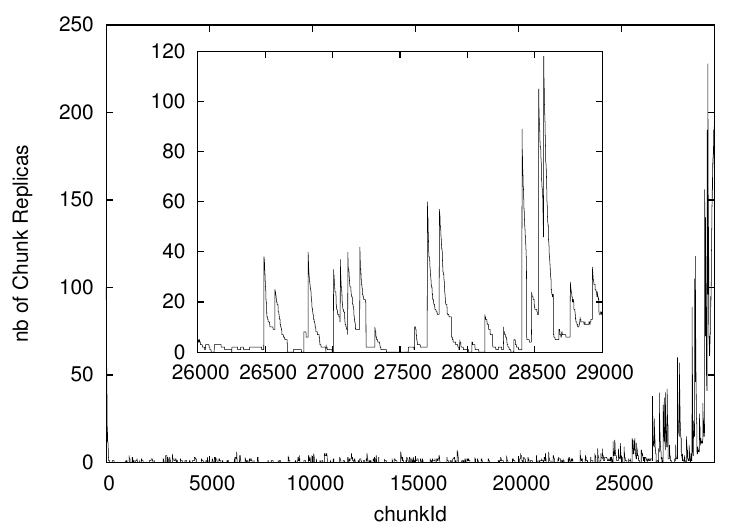}
}
\subfigure[Turntable]{\label{fig:pieceReplicasForTurntable}
\includegraphics[width=0.48\columnwidth]{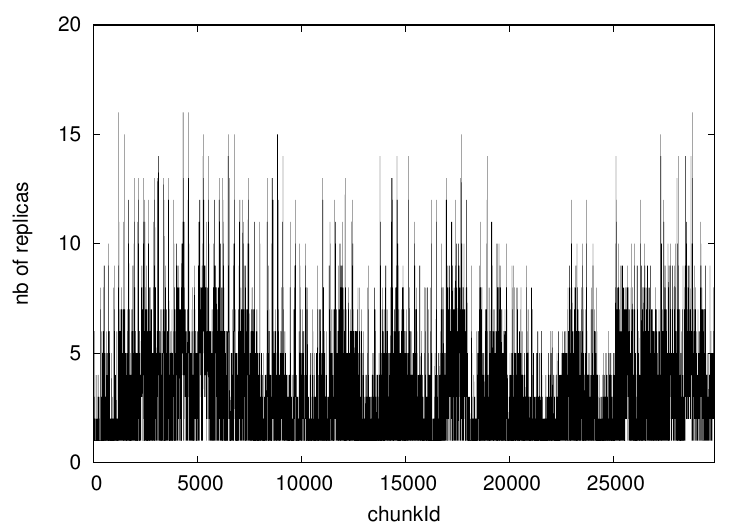}
}
\subfigure[P2TSS-Rand]{\label{fig:pieceReplicasForRandomP2TSS}
\includegraphics[width=0.48\columnwidth]{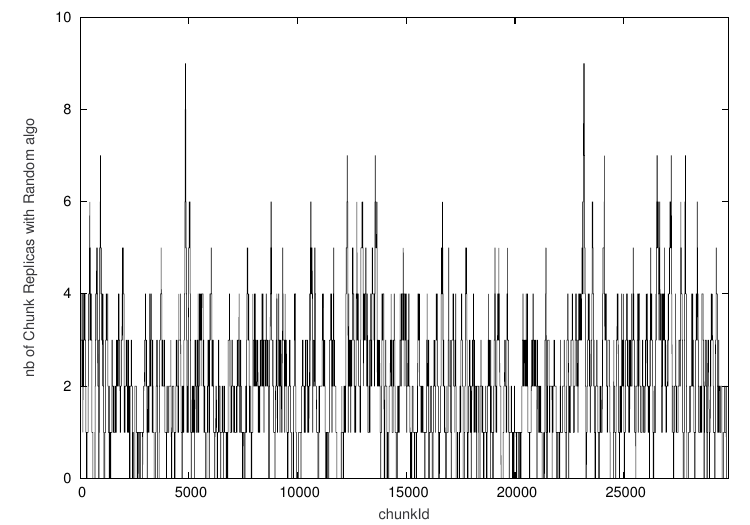}
}
\subfigure[P2TSS-Live]{\label{fig:pieceReplicasForLiveP2TSS}
\includegraphics[width=0.48\columnwidth]{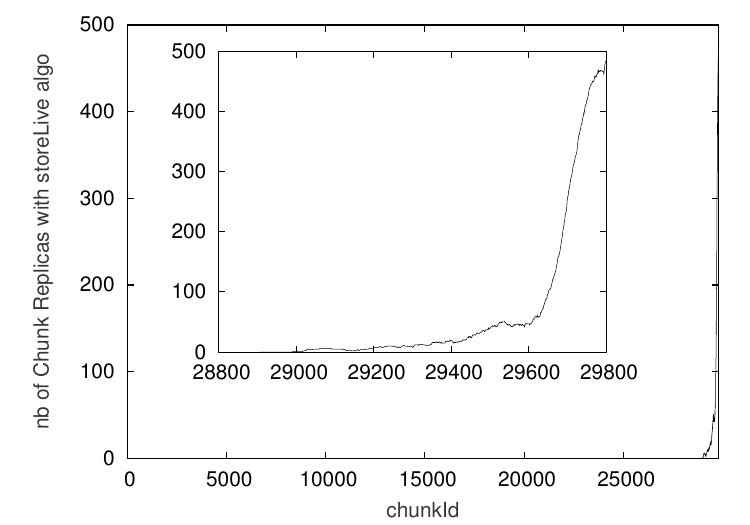}
}
\caption{Number of replicas for every piece at the end of the simulation}\label{piece-replicas}
\end{figure*}


We present in Figure~\ref{piece-replicas} the evolution of the number
of replicas of every piece from time 0 to time $30,\!000$ at the end
of our simulation for four systems: respectively PACUS, Turntable,
P2TSS with the initial playback caching (hereafter called P2TSS-Rand)
and finally P2TSS with the live stream playback caching (called
P2TSS-Live).

We first analyse the results obtained by PACUS and its cache-and-relay
approach.  In the embedded figure~\ref{fig:pieceReplicasForPacus}, we
zoom on a smaller area from time $26,\!000$ to time $29,\!000$.  This
result shows that the most replicated pieces are the latest pieces.
Now, we invite the reader to recall Figure~\ref{fig:playingposition}
where the lags of shifters are represented.  The number of piece
replicas in Figure~\ref{fig:pieceReplicasForPacus} matches piece
popularity in Figure~\ref{fig:playingposition}. The more requested are
the pieces, the more replicated they are. Even the variable popularity
of programs produce differences between piece replicas. We highlight
here that cache-and-relay approach (and peer-assisted architecture)
fits remarkably with time-shifted TV: the storage capacity of nodes is
automatically utilized for the most popular pieces.  Please note
however that some pieces in early period do not have any replica.  It
means that a viewers requesting this period has no other choice but to
request missing pieces from the server.

Results for the Turntable structure are in
Figure~\ref{fig:pieceReplicasForTurntable}.  The variation of the
number of replicas is one order of magnitude smaller than in PACUS,
so, as we previously shown, some requests cannot be fulfilled because
nodes are congested. However, every piece has at least one replica in
the system despite node churn. Therefore, the presence of pieces is
guaranteed in the turntable system, even for a long service duration.

Finally, we describe the results for both P2TSS implementations, that
is P2TSS-Rand (see Figure~\ref{fig:pieceReplicasForRandomP2TSS}) and
P2TSS-Live (see Figure~\ref{fig:pieceReplicasForLiveP2TSS}).  The
random algorithm of P2TSS-Rand achieves an almost uniform piece
distribution, however it does not guarantee that all pieces are
actually replicated at least once in the system. Besides, this
algorithm suffers from the same potential drawbacks as Turntable on
the most requested pieces that are insufficiently
replicated. Contrarily, the live piece algorithm of P2TSS-Live assigns
the system storage for the most recent pieces and live piece. Here,
the availability of old pieces is not guaranteed at all because nodes
do not stay in the system during the whole simulation, which is a
realistic case. It results that about $3\%$ of pieces are replicated
in P2TSS-Live.

\paragraph*{Gain at the server side}

\begin{table}
\centering
\begin{tabular}{cc|cccccc}
  \multicolumn{2}{c}{\emph{}}
    & \multicolumn{2}{|c}{\emph{from nodes}}
    & \multicolumn{4}{|c} {\emph{from server}} \\
  \multicolumn{2}{c}{\emph{}} & \multicolumn{2}{|c}{\emph{}} &
  \multicolumn{2}{|c}{\emph{missing piece}}
    &\multicolumn{2}{c}{\emph{no capacity}} \\ \hline \hline
  \multicolumn{2}{c}{\emph{PACUS}}
    & \multicolumn{2}{|c}{75.2\%}
    & \multicolumn{2}{|c}{0.1\%}
    & \multicolumn{2}{|c}{24.7\%} \\ \hline
  \multicolumn{2}{c}{\emph{Turntable}}
    & \multicolumn{2}{|c}{78.5\%}
    & \multicolumn{2}{|c}{0\%}
    & \multicolumn{2}{|c}{21.5\%} \\ \hline
  \multicolumn{2}{c}{\emph{P2TSS-Rand}}
    & \multicolumn{2}{|c}{11.2\%}  &\multicolumn{2}{|c}{23\%}
    & \multicolumn{2}{|c}{65.8\%} 
    \\ \hline
  \multicolumn{2}{c}{\emph{P2TSS-Live}}
    & \multicolumn{2}{|c}{2.8\%}  &\multicolumn{2} {|c}{22.8\%}
    & \multicolumn{2}{|c}{74.4\%}
    \\ \hline

\end{tabular}
\caption{Ratio of pieces from nodes vs. failure ratio} \label{tab:capacity}
\end{table}

We now present in Table~\ref{tab:capacity} the percentage of piece
requests that are handled by nodes (first column). We also analyze the
reasons that explain why a request has not been treated by a node. In
the second column, we reveal the number of requests for pieces that do
not contain any replica in the system. In the third colum, we show the
number of times the piece is replicated, but no node can serve the
request.  
The percentage of piece requests handled by nodes indicates also the
benefits that a service provider can expect from an implementation of
a decentralized architecture.  Obviously, the larger is the percentage,
the better is the system because the less saturated is the server.
PACUS and Turntable obtains satisfying results as more than three
quarters of the requests are handled by nodes.  On the contrary, both
P2TSS implementations obtain low results: less than $12\%$ of 
requests are treated by the nodes.

As can be expected, the percentage of missing piece for Turntable is
equal to 0\%. But, to our surprise, the number of missing piece is
almost equivalent for PACUS. Actually, all recent measurements have
shown that the behavior of clients is reproductible, therefore, for a
request for one piece, the probability that another node has requested
it recently is close to 1. The pieces that do not exist any longer in
the system are actually never requested.

Let finally have a closer look at the results of P2TSS. For
P2TSS-Rand, the random algorithm makes that, unforunately, the most
recent pieces, which are also the most requested, are not stored in
the shared buffer of nodes. Therefore, 35\% of the requests do not
find any node storing the requested piece. And the replication number
of the recent pieces is also too low for the generated traffic. This
result prove that a static replication management is irrelevant. For
P2TSS-Live, the most recent pieces are stored, but, as shown in the
measurements, the majority of requests are for pieces that are more
than three hours old, although the time-to-live of clients is often
smaller. Therefore, most of the requested pieces are not
available. This result proves that storing the most recent pieces is
not a good strategy because too many peers store the same, not so
requested, pieces.


\section{Conclusion}  \label{sec:conclu}

Time shifted TV is a new service whose implementation is challenging
although it represents for the TV broadcasters a critical
transformation of their model. TV broadcasters need large-scale
delivery systems for their streams, unfortunately previous works have
only sketched unconvincing solutions. This paper partly addresses this
issue by exploring solutions based on decentralized architectures. For
the PACUS system, we demonstrate that a lightweight and self-adaptive
peer-assisted architecture can absorb most of the traffic.  The
Turntable structure is, to our knowledge, the first fully distributed
structure that ensures availability of video pieces. Both proposals
are motivated by a context, PACUS for large TV broadcasters and
turntable for small independent channels.

This paper is a first step toward the understanding of the whole
complexity of delivering time-shifted TV streaming at
large-scale. Recent efforts from measurement institutes will soon
offer some more precise insights about the behavior of clients. Some
of the assumption we made in this paper may be challenged by the new
usages of this service. The server vendors are also designing
next-generation of routers, which will be able to both ingest and
deliver content more efficiently, these efforts however being
counterbalanced by the growing popularity of the service.

Our plans for future works include the development of a partnership
with a popular independent Internet TV provider, which will allow us
to experiment in the real world our proposals. We will also design
algorithms that leverage on Dynamic Adaptive Streaming over HTTP and
scalable video coding to adapt the quality of the video during peaks.


\bibliographystyle{IEEEtran}

\bibliography{references}

\end{document}